# Interacting Floquet topological magnons in laser-irradiated Heisenberg honeycomb ferromagnets

Hongchao Shi[a], Heng Zhu[a], Bing Tang[*], Chao Yang

*Department of Physics, Jishou University, Jishou 416000, China*

**ABSTRACT**

When a Heisenberg honeycomb ferromagnet is irradiated by high-frequency circularly polarized light, the underlying uncharged magnons acquire a time-dependent Aharonov–Casher phase, which makes it a Floquet topological magnon insulator. In this context, we investigate the many-body interaction effects of Floquet magnons in laser-irradiated Heisenberg honeycomb ferromagnets with ocontaining Dzyaloshinskii-Moriya interaction under the application of circularly polarized off-resonant light. We demonstrate that the quantum ferromagnet systems periodically laser-driven exhibits temperature-driven topological phase transitions due to Floquet magnon-magnon interactions. The thermal Hall effect of Floquet magnons serves as a prominent signature for detecting these many-body effects near the critical point, enabling experimental investigation into this phenomenon. Our study complements the lack of previous theoretical works that the topological phase transition of the Floquet magnon under the linear spin wave approximation is only tunable by the light field. Our study presents a novel approach for constructing Floquet topological phases in periodically driven quantum magnet systems that goes beyond the limitations of the linear spin wave theory. We provide numerical results based on the well-known van der Waals quantum magnet $CrX_3$ (X=F, Cl, Br, and I), calling for experimental implementation.

## I. INTRODUCTION

[a]These authors contributed equally to this work
[*] bingtangphy@jsu.edu.cn

Over the past few decades, studies of the topological insulators and topological phase have made great progress in the field of condensed matter physics [1-8]. In analogy to electronic systems, the topological phases have also been extended to bosonic systems, such as photonic[9-11], phononic [12,13], and magnonic systems [14-17]. There is a surge of interest in utilizing magnons, a low-energy collective excitation in magnets [18,19] that are easily manipulated by magnetic fields, have low-dissipation and permit a pure spin transport without Joule heating, for spintronics [20,21].

Recently, the thermal magnon Hall effect has been realized experimentally in the insulating quantum kagome ferromagnets Cu(1-3, bdc) [22,23] and the pyrochlore ferromagnets $Lu_2V_2O_7$, $Ho_2V_2O_7$, $In_2Mn_2O_7$ [24,25] following a theoretical proposal [26,27]. It is generally believed that thermal magnon Hall effect results from the nontrivial topology of magnon dispersions [26-31] encoded in the Berry curvature induced by the Dzyaloshinskii-Moriya interaction (DMI) [32, 33], which plays the role of spin orbit coupling. In insulating quantum magnets the DMI is an intrinsic anisotropy and it is present due to the lack of inversion symmetry of the lattice. For honeycomb magnets, the midpoint between two magnetic ions on the next-nearest neighbour bonds is not an inversion center. Therefore, a DMI is allowed on the honeycomb lattice and a magnon analogue of the Haldane model [1] can be realized in honeycomb ferromagnets[34-37]. Thus, the thermal magnon Hall effect also to exists in honeycomb magnets. It is worth noting that the early study of topological magnons mainly based on the linear spin wave theory, where the interactions between magnons can be safely ignored [26-31, 34-36].

But in actuality, as the temperature increases, the influence of magnon-magnon interactions becomes more pronounced, which are typically treated as small terms that often expanded using techniques such as Holstein-Primakoff (HP) [38] or Dyson-Maleev transformations [39,40]. Recently, the importance of the many-body interactions effect has been recognized in magnonic systems, where the interactions between magnons lead to magnon decays and spectral renormalizations [41-50]. These scattering processes can introduce intriguing momentum or temperature-dependent behaviors of magnons in magnet systems, which have not been previously explored in the study conducted using the linear spin-wave theory [26-31,34-36]. In particular, there have been some investigations on the effects of interactions on the Dirac magnons in a honeycomb lattice [45-48] show that the phase transition of magnons at a critical temperature is driven by their interactions, where the DMI is essential in determining the topological properties of magnon-magnon interaction in honeycomb ferromagnets.

Also, laser-irradiation of solid-state materials has attracted considerable attention and interest as an alternative way for engineering topological nontrivial states from topologically trivial quantum materials recently [51-56]. In this formalism, topologically trivial systems can be periodically driven to nontrivial topological systems termed Floquet topological insulators [53,54]. They have an advantage over their static (equilibrium) topological counterpart, in that their intrinsic properties can be manipulated and different topological phases can be achieved. In irradiated insulating quantum magnets with charge-neutral magnons [57-60], the Floquet physics can emerge from the coupling of the electron spin magnetic dipole moment to the laser electric field through the time-dependent version of the static Aharonov-Casher phase [61], which acts as a vector potential or gauge field to the spin current. A quantum magnet system driven periodically can be studied by the Floquet-Bloch theory [62]. Previously it has been shown that a tunable DMI by laser field in a two-dimensional (2D) laser-irradiated Heisenberg ferromagnets can induce photoinduced topological phase transition[57,60]. However, to the best of our knowledge, the topological property of the Floquet magnon with many-body interactions effects at finite temperatures involved has not been studied yet. How the magnon interactions affect the topology of laser-irradiated insulating quantum magnets is still an issue of fundamental interest in topological quantum materials

In this work, we investigate the many-body interaction effects of Floquet magnons in a laser-irradiated Heisenberg honeycomb ferromagnet with off-resonant circularly polarized light field tunable DMI. Using linear spin wave and magnonic Floquet-Bloch theory, we show that when the magnet systems are periodically driven by off-resonant circularly polarized lights, they effectively map onto the corresponding static spin model plus a tunable photoinduced magnetic field along the $\hat{\mathbf{z}}$ direction, which is perpendicular to the honeycomb plane. In particular, when the many-body interaction effects of Floquet magnons are considered, we combine magnonic Floquet-Bloch theory and Green's function method, we find that the topological phase transitions in laser-irradiated Heisenberg honeycomb ferromagnets are driven by temperature due to the Floquet magnon-magnon interactions. These transitions are marked by the Floquet magnon band gap closing-reopening with increasing temperature and the sign change of the Chern number and thermal Hall conductivity with increasing temperature at a finite off-resonant circularly polarized light field and magnetic field.

Our paper is organized as follows. In Sec. II, we sketch the model of this work. In Sec. III, we presented the magnonic Floquet-Bloch theory and Green's function method. In Sec. IV, we have shown the magnon band structure of Heisenberg honeycomb ferromagnet under linear spin wave theory and consider the magnon-magnon interaction, respectively. In Sec. V, we discuss the topological properties of interacting Floquet magnons. In Sec. VI, We investigate the thermal transport property. The summary and outlook are given in Sec. VII.

## II. MODEL

In this article, we take into account a two-dimensional (2D) laser-irradiated Heisenberg honeycomb ferromagnet, whose lattice structure is shown in Fig. 1. An external magnetic field is also considered. This ferromagnetic system is described by the following spin Hamiltonian

$$\mathcal{H} = -J\sum_{\langle i,j \rangle} \mathbf{S}_i \cdot \mathbf{S}_j + \sum_{\langle\langle i,j \rangle\rangle} \mathbf{D}_{ij} \cdot \left( \mathbf{S}_i \times \mathbf{S}_j \right) - g\mu_B \sum_i \mathbf{B} \cdot \mathbf{S}_i, \tag{1}$$

where $\langle i,j \rangle$ and $\langle\langle i,j \rangle\rangle$ represent the summation over the nearest and next-nearest neighbor sites, respectively. $J$ and $D$ represent the pairwise interactions between the nearest neighboring ions and the DMI between the next-nearest neighboring ions, respectively. The DMI vector is parallel to the z-direction with $\mathbf{D}_{ij} = v_{ij} D \hat{\mathbf{z}}$, where $v_{ij} = \pm 1$ for counter-clockwise and clockwise hopping spin magnetic moments on each honeycomb layer sublattice. The applied magnetic field along the z axis with $\mathbf{B} = B \cdot \hat{\mathbf{z}}$, and we define $h = g \mu_B B$ being the magnetic field strength, g is the spin $g$-factor and $\mu_B$ is the Bohr magneton.

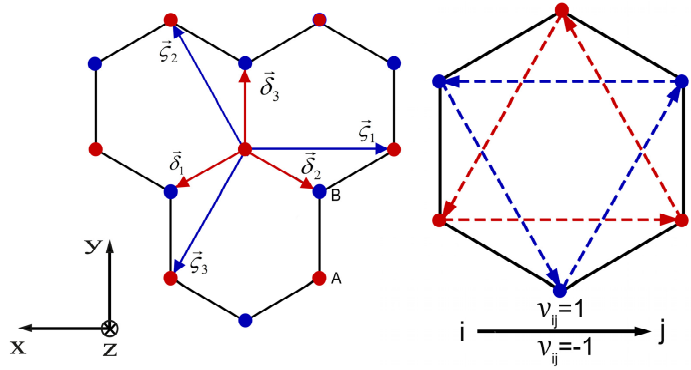

**Fig. 1.** (Color online) Schematic of the honeycomb lattice structure, which is made up of two triangular sublattices.

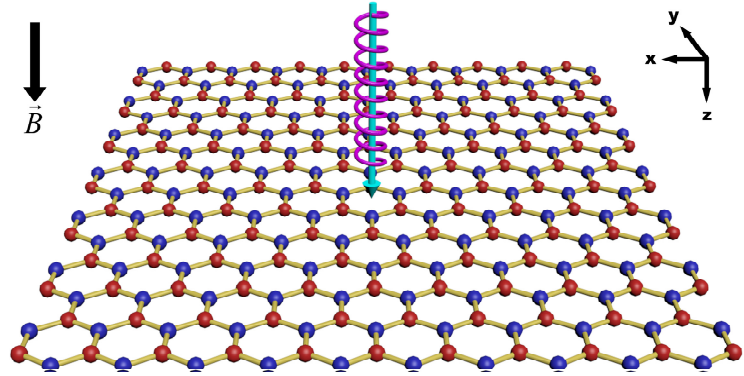

**Fig. 2.** (Color online) Schematic representation of a honeycomb ferromagnet being illuminated by a circularly polarized laser (perpendicular to the ferromagnet plane).

We consider a circularly polarized laser irradiated onto the 2D Heisenberg honeycomb ferromagnet, as shown in Fig. 2. Physically, charge-neutral magnons can interact with an electromagnetic field through their spin magnetic dipole moment. The corresponding time-dependent version of the Aharonov-Casher phase emerges explicitly from quantum field theory with the Dirac-Pauli Lagrangian. We take the spin magnetic dipole moment carried by magnons to be along the z-direction $\mu = g\mu_B e_z$, where $g$ is called the $g$-factor and $\mu_B = e\hbar/2m_e$ is known as the Bohr magneton. In the presence of a laser (electric) field $E(t)$, the hopping spin magnetic dipole moments accumulate a time-dependent version of the Aharonov-Casher phase [61]

$$\phi_{ij}(t) = \frac{g\mu_B}{\hbar c^2} \int_{r_i}^{r_j} \Xi(t) \cdot dl. \tag{2}$$

where $\hbar$ is the reduced Planck's constant, and $c$ is the speed of light. $\Xi(t) = \mathbf{E}(t) \times \hat{\mathbf{z}}$ with $\mathbf{E}(t) = -\partial_t \mathbf{A}(t)$, $\mathbf{A}(t)$ is the time-dependent vector potential given by

$$\mathbf{A}(t) = A_0 \left[ \sin(\omega t), \cos(\omega t), 0 \right] \tag{3}$$

where $A_0 = E_0/\omega$ is the strength of the time-dependent vector potential. $E_0$ stands for the amplitude of the electric field, $\omega$ represents the circular frequency of the light wave. The corresponding time-dependent oscillating electric field is given by

$$\Xi(t) = E_0 \left[ \sin(\omega t), \cos(\omega t), 0 \right]. \tag{4}$$

The resulting time-dependent Hamiltonian is given by

$$\mathcal{H}(t) = -J\sum_{\langle i,j \rangle} \left[ S_i^z S_j^z + \frac{1}{2}\left( e^{-i\phi_{ij}(t)} S_i^- S_j^+ + H.c. \right) \right] - \frac{D}{2} \sum_{\langle\langle i,j \rangle\rangle} \left( iv_{ij} e^{-i\phi'_{ij}(t)} S_i^- S_j^+ + H.c. \right) - h\sum_i S_i^z. \tag{5}$$

Noticing that the direction of the vector pointing from $i$ to $j$ defines a relative angle $\phi_{ij}$ and $\phi'_{ij}$, we get $\phi_{ij}(t) = \varepsilon_0 \sin(\omega t - \phi_{ij})$ and $\phi'_{ij}(t) = \varepsilon'_0 \sin(\omega t - \phi'_{ij})$, $\frac{g\mu_B}{\hbar c^2}$ is absorbed in $\varepsilon_0$ and $\varepsilon'_0 = \sqrt{3}\varepsilon_0$. We focus on the linear spin-wave approximation, which is reasonable in the large spin value limit and the low-temperature regime. This can be implemented via recasting the spin operators in the time-dependent Hamiltonian in terms of the following linearized HP transformation [38]

$$S_i^+ = \sqrt{2S} a_i, \quad S_i^- = \sqrt{2S} a_i^\dagger, \quad S_i^z = S - a_i^\dagger a_i. \tag{6}$$

Here, $a_i^\dagger (a_i)$ is the magnon creation (annihilation) operator. The resulting linear bosonic Hamiltonian has time periodicity, i.e., $\mathcal{H}(t+T)=\mathcal{H}(t)$, where $T=\dfrac{2\pi}{\omega}$ corresponds to the period of the laser field.

### III. METHODS
### A. Magnonic Floquet-Bloch theory

The Floquet-Bloch theory[62] is a formalism for studying periodically driven quantum systems and it applies to different cases of physical interests. The magnonic version describes the interaction of light with magnonic Bloch states in insulating quantum magnets. In the present case, the time-dependent Hamiltonian $\mathcal{H}(\mathbf{k},t)$ can be obtained by making the time-dependent Peierls substitution $\mathbf{k} \rightarrow \mathbf{k}+\mathbf{A}(t)$. Note that the $\mathcal{H}(\mathbf{k},t)$ is periodic due to the time-periodicity of the vector potential. Hence, it can be expanded in Fourier space as

$$\mathcal{H}(\mathbf{k},t)=\mathcal{H}(\mathbf{k},t+T)=\sum_{n=-\infty}^{\infty} e^{in\omega t}\mathcal{H}_n(\mathbf{k}), \qquad (7)$$

where $\mathcal{H}_n(\mathbf{k})=\dfrac{1}{T}\int_0^T e^{-in\omega t}\mathcal{H}_n(\mathbf{k},t)dt = \mathcal{H}_{-n}^\dagger(\mathbf{k})$ is the Fourier component. Therefore, its eigenvectors in the Floquet-Bloch theory can be written as $\mathcal{H}(\mathbf{k},t)|\psi_\alpha(\mathbf{k},t)\rangle = e^{-i\omega\varepsilon_\alpha(\mathbf{k})t}|\xi_\alpha(\mathbf{k},t)\rangle$, $|\xi_\alpha(\mathbf{k},t)\rangle = |\xi_\alpha(\mathbf{k},t+T)\rangle = \sum_n e^{in\omega t}|\xi_\alpha^n(\mathbf{k})\rangle$ is the time-periodic Floquet-Bloch wave function of magnons and $\varepsilon_\alpha(\mathbf{k})$ are the magnon quasi-energies. We define the Floquet operator as $\mathcal{H}^F(\mathbf{k},t)=\mathcal{H}(\mathbf{k},t)-i\partial_t$, which leads to the Floquet eigenvalue equation

$$\sum_m \left[\mathcal{H}^{n-m}(\mathbf{k})+m\omega\delta_{n,m}\right]\xi_\alpha^m(\mathbf{k})=\varepsilon_\alpha(\mathbf{k})\xi_\alpha^n(\mathbf{k}). \qquad (8)$$

### B. Green's function

In this section, we will introduce the perturbation methods of the many-body Green's function[62] in order to study the many-body effect of Floquet magnons and its interplay with thermal fluctuation in later study. We define a matrix Green's function as $\mathcal{G}(\mathbf{k},\tau) = -\langle \mathcal{T}_\tau \psi_\mathbf{k}(\tau) \psi_\mathbf{k}^\dagger(0) \rangle$, where $\mathcal{T}_\tau$ is a time-ordering operator for the imaginary time $\tau = it$, and $0 \leq \tau \leq \beta$ with $\beta = (k_B T)^{-1}$. The $\tau$-dependent operator is defined as $O(\tau) = e^{\mathcal{H}\tau} O(0) e^{-\mathcal{H}\tau}$, which is formally obtained by the analytic continuation to imaginary time $\tau$ of the Heisenberg operator $O(t)$, and $\mathcal{H} = \mathcal{H}_0 + \mathcal{H}_{\text{int}}$ is the zeroth-order static time-independent effective Floquet Hamiltonian. The bracket $\langle \cdots \rangle$ denotes the thermodynamic average. The first-order Hartree Feynman diagrams resulting from Floquet magnon-magnon interactions are depicted in Fig. 2.

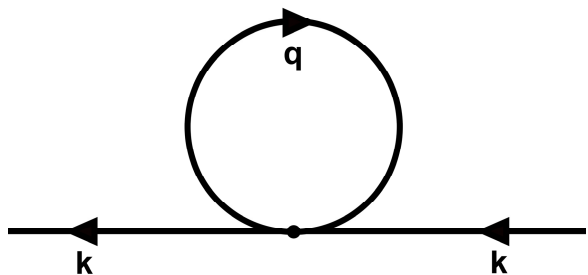

Fig3. The Feynman diagram of the Hartree contributes to $1/S$ many-body corrections in linear spin-wave theory. The solid line represents a Floquet magnon and the arrow denotes the propagation direction. With number-conserving four-Floquet magnon vertex is indicated by a black circle. Variable $q$ denotes the momentum of thermally excited Floquet magnons.

Based on the first-order Hartree diagram shown in Fig. 3, to get the solution, we solve the Heisenberg equation of motion for the Green's function elements and apply the random phase approximation to extract the nonlinear self-energy corrections from Floquet magnon-magnon interactions. After a Fourier transformation

$$\mathcal{G}(\mathbf{k},\tau) = \frac{1}{\beta} \sum_n e^{-i\omega_n \tau} \mathcal{G}(\mathbf{k},\omega_n), \tag{9}$$

where $\omega_n$ is the bosonic Matsubara frequency. The Green's functions $\mathcal{G}(\mathbf{k},\omega_n)$

satisfy a matrix Dyson's equation $\mathcal{G}^{-1}(\mathbf{k},\omega_n) = i\omega_n - H_1(\mathbf{k})$, where $\mathcal{G}^{-1}$ is the inverse of the Green's function, and $H_1(\mathbf{k})$ is the renormalized effective Hamiltonian.

### IV. MAGNONIC FLOQUET BAND

*(i). Linear Hamiltonian for Laser-Irradiated Heisenberg honeycomb ferromagnets*

In order to study the periodically laser driven Heisenberg honeycomb ferromagnet in Eq. (4), we will apply the Floquet theory to transform the present time-dependent spin model to one static time-independent effective spin model, which is describe by the Floquet Hamiltonian. Physically, this static effective Hamiltonian $\mathcal{H}_{eff}$ can be expressed in terms of $\omega^{-1}$, namely, $\mathcal{H}_{eff} = \sum_{m \geq 0} \omega^{-m} \mathcal{H}^m$, where $m$ is an integer. When the circular frequency $\omega$ of the laser is much larger than the magnon frequency bandwidth $\Delta$, i.e., $\omega \gg \Delta$, this way is applicable. In the current work, we pay attention to the off-resonant regime, thus it suffices to take into account the zeroth order of the static Floquet Hamiltonian [52,56,59,60]. By making use of the discrete Fourier component of the time-dependent Hamiltonian, we can obtain a zeroth order effective Hamiltonian $\mathcal{H}_0 = \mathcal{H}^0$, where $\mathcal{H}^m = \frac{1}{T}\int_0^T dt e^{-im\omega t} \mathcal{H}(t)$. In the momentum space, the effective magnon Hamiltonian can be written as $\mathcal{H}_0 = \sum_{\vec{k}} \psi_\mathbf{k}^\dagger H_0(\mathbf{k}) \psi_\mathbf{k}$, where $\psi_\mathbf{k}^\dagger = (a_\mathbf{k}^\dagger, b_\mathbf{k}^\dagger)$ is a two-component spinor operator and the Bogoliubov Hamiltonian is given by

$$H_0(\mathbf{k}) = h_0 \sigma_0 + h_x(\mathbf{k})\sigma_x + h_y(\mathbf{k})\sigma_y + h_z(\mathbf{k})\sigma_z. \qquad (10)$$

where $h_0 = 3JS + h$, $h_x(\mathbf{k}) = -J_F S \operatorname{Re}(\gamma_\mathbf{k})$, $h_y(\mathbf{k}) = J_F S \operatorname{Im}(\gamma_\mathbf{k})$, $h_z(\mathbf{k}) = 2D_F S \rho_\mathbf{k}$, $\gamma_{\vec{k}} = \sum_{n=1}^3 e^{i\vec{k}\cdot\vec{\delta}_n}$, $\rho_\mathbf{k} = \sum_{n=1}^3 \sin(\mathbf{k}\cdot\zeta_n)$, and the vectors $\delta_n$ and $\zeta_n$ are shown in Fig.1. $J_F = JJ_0(\varepsilon_0)$, and $D_F = DJ_0(\varepsilon_0')$, where $J_m(x)$ is the Bessel function of order

$m \in Z$. $\sigma_i (i=x,y,z)$ are Pauli matrices and $\sigma_0$ is identity matrix. In fact, the light intensity of the laser can be characterized via a dimensionless quantity $\varepsilon_0 = g\mu_B E_0 a / \hbar c^2$. The Floquet magnon bands are given by

$$\varepsilon_\alpha^0 (\mathbf{k}) = h_0 + \lambda \varepsilon(\mathbf{k}), \qquad (11)$$

where $\varepsilon(\mathbf{k}) = \sqrt{h_x(\mathbf{k})^2 + h_y(\mathbf{k})^2 + h_z(\mathbf{k})^2}$ and $\lambda = 1(-1)$ corresponds to the up(down) band, namely, $\alpha = u(\alpha = d)$.

The tunable Floquet magnon energy bands and Density of states are depicted in Fig. 4(a) and Fig. 4(b), respectively. The results show that the magnon band and the Density of states can be tuned by the circularly polarized light field.

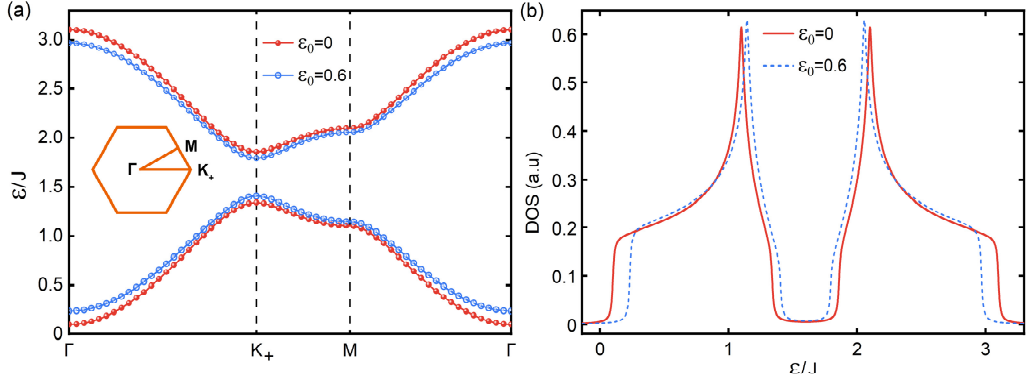

**Fig. 4.** (Color online) (a)The tunable Floquet magnon energy bands along the high-symmetry lines $\Gamma - K_+ - M - \Gamma$. (b) The tunable Floquet Density of states per unit cell of the ferromagnet on the honeycomb lattice. The other parameters are set to $D = 0.1J$, $h = 0.1J$, and $S = 1/2$.

In fact, the laser field tunable DMI can cause an tunable effective Haldane mass term, which will open up a tunable non-trivial band gap $\Delta_{K_+} = 6\sqrt{3} D_F S$ between the upper and lower branches at the Dirac points $K_+$. In the absence of the laser field, the size of the band gap at the Dirac point is $\Delta'_{K_+} = 6\sqrt{3} DS$. As the laser field is applied, the zero energy mode is lifted for $\varphi = \pi/2$, which implies a photoinduced magnetic order with-out a high applied magnetic field [64]. In the presence of laser fields, the magnonic excitation energy at $\Gamma$ is given by the tunable photoinduced magnetic field and the magnetic field $h$.

*(ii). The renormalized magnon band for Interacting topological Dirac magnons*

We relate the spin and boson operators using the HP transformation truncated to the first order in 1/S as follows,

$$S_i^+ = \sqrt{2S}\left(a_i - \frac{a_i^\dagger a_i a_i}{4S}\right), \quad S_i^- = \sqrt{2S}\left(a_i^\dagger - \frac{a_i^\dagger a_i^\dagger a_i}{4S}\right), \quad S_i^z = S - a_i^\dagger a_i. \tag{12}$$

The higher-order terms from the HP transformation in Eq. (8) give the zeroth-order static time-independent effective interaction Hamiltonian associated with Eq.(4)

$$\mathcal{H}_{int} = \frac{J_F}{4N}\sum_{\{k_i\}}\left(\gamma_{k_4}^* b_{k_1}^\dagger b_{k_2}^\dagger b_{k_3} a_{k_4} + \gamma_{k_2} b_{k_1}^\dagger a_{k_2}^\dagger b_{k_3} b_{k_4} + \gamma_{k_4} a_{k_1}^\dagger a_{k_2}^\dagger a_{k_3} b_{k_4} + \gamma_{k_2}^* a_{k_1}^\dagger b_{k_2}^\dagger a_{k_3} a_{k_4}\right)$$
$$-\frac{J}{4N}\sum_{\{k_i\}}\left(4\gamma_{k_4-k_2} a_{k_1}^\dagger b_{k_2}^\dagger a_{k_3} b_{k_4}\right) - \frac{D_F}{2N}\sum_{\{k_i\}}\left(\rho_{k_2} + \rho_{k_4}\right)\left(a_{k_1}^\dagger a_{k_2}^\dagger a_{k_3} a_{k_4} - b_{k_1}^\dagger b_{k_2}^\dagger b_{k_3} b_{k_4}\right) \tag{13}$$

where $\{k_i\}$ stands for summation over all $k_i$. In Eq. (8), the conservation of momentum is due to $\frac{1}{N}\sum_i e^{i(k_1+k_2-k_3-k_4)\cdot\vec{r}_i} = \delta_{k_1+k_2,k_3+k_4}$. Similar to a recent new study[47], these interaction terms are able to be rewritten as a more compact expression such as $\mathcal{H}_{int} = \sum_{\{k_i\}} V_{k_3,k_4}^{k_1,k_2} \psi_{k_1}^\dagger \psi_{k_2}^\dagger \psi_{k_3} \psi_{k_4}$. We can get the effective Hamiltonian by using the Green's function method which is introduced in Sec. III. B, written as

$$H_1(\mathbf{k}) = H_0(\mathbf{k}) + \sum\nolimits^{(1)}(\mathbf{k}) = \begin{pmatrix} h_0 + p_q + m_k & -(J_F S - M)\gamma_k - v_k \\ -(J_F S - M)\gamma_k^* - v_k^* & h_0 + p_q - m_k \end{pmatrix}, \tag{14}$$

where $p_q = \frac{1}{2N}\sum_q\left(J_F|\gamma_q|\sqrt{1-\chi_q^2}v_q^- - J\gamma_0 v_q^+\right) + \frac{D_F}{N}\sum_q \rho_q \chi_q v_q^-$, $m_k = 2D_F S\rho_k - \delta_k$,

$M = \frac{J_F}{2N}\sum_q v_q^+$, $\delta_k = \frac{D_F \rho_k}{N}\sum_q v_q^+$, $v_q^+ = f(\varepsilon_d^0(\mathbf{q})) + f(\varepsilon_u^0(\mathbf{q}))$, $v_q^- = f(\varepsilon_d^0(\mathbf{q})) - f(\varepsilon_u^0(\mathbf{q}))$,

and $v_k = \frac{J}{2N}\sum_q \gamma_{k-q} e^{i\phi_q}\sqrt{1-\chi_q^2}v_q^-$. Here, $\beta = 1/k_B T$, $f(\varepsilon_\alpha^0(\mathbf{q})) = \left[e^{\beta\varepsilon_\alpha^0(\mathbf{q})} - 1\right]^{-1}$ is

the Bose-Einstein distribution function, and $\chi_{\mathbf{q}} = h_z(\mathbf{q})/\varepsilon(\mathbf{q})$.

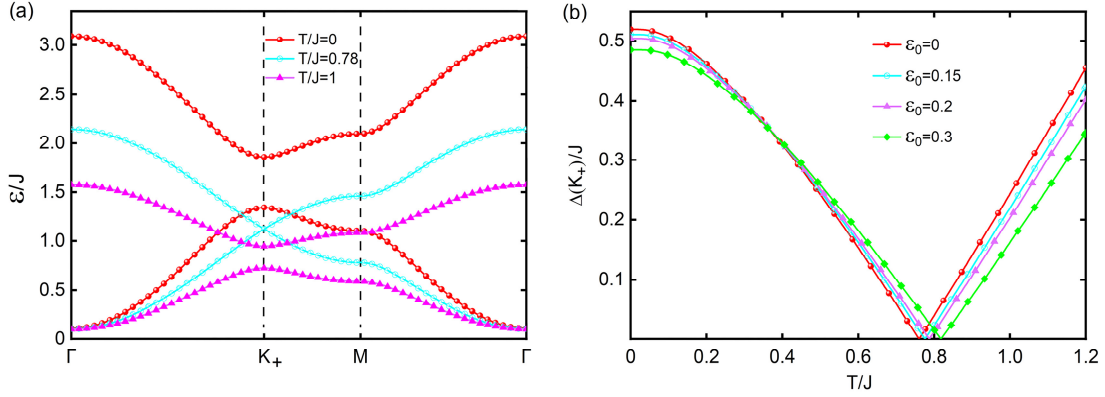

**Fig. 5.** (Color online) The renormalized Floquet magnons band structures driven by periodically circularly polarized light $\varphi = \pi/2$. (a) The renormalized Floquet magnon dispersion curves along the high-symmetry lines $\Gamma - K_+ - M - \Gamma$, and the light intensity $\varepsilon_0 = 0.15$. (b) The gap $\Delta$ at the Dirac point $K_+$ as a function of temperature. The parameters are chosen as $D = 0.1J$, $h = 0.1J$, and $S = 1/2$.

In Fig. 5(a), we show the renormalized Floquet magnon bands of periodically circularly polarized light-driven honeycomb ferromagnets at three different temperatures. It is clearly seen that, with the increase of the temperature, the renormalized Floquet magnon band gap at the Dirac points $K_+$ decreases and will close at approximately $T_c = 0.78J$. If the temperature is further elevated, the renormalized Floquet magnon band gap reopens at the Dirac point $K_+$ and its width increases with $T$. To understand the influence of the light intensity on width of the band gap, we plot band gaps at the Dirac point $K_+$ as a function of temperature corresponding to different the light intensity $\varepsilon_0$, as shown in Fig. 5(b). It can be clearly seen that the critical temperature $T_c$ for the gap-closing increases as increasing the light intensity $\varepsilon_0$. The temperature can derive a gap-closing phenomenon, which means a topological phase transition may occur with the increase of the temperature, and we will discuss it later.

## V. TOPOLOGICAL PROPERTIES OF INTERACTING FLOQUET MAGNON

The Berry curvature is one of the main important quantities in topological systems, which provides an effective gauge field for the bosonic magnons in the k-space and dominates the transport properties. In Floquet topological systems, it is customary to assume that the quasienergy levels of the Floquet Hamiltonian are close to the equilibrium system, which is realized in the off-resonant limit [52,59,65]. Therefore, the properties of equilibrium topological systems can be applied to Floquet topological systems. To investigate the magnon transport in Floquet topological systems, we define the Berry curvature of the Floquet magnon bands as

$$\Omega_\alpha^F(\mathbf{k}) = i\nabla_\mathbf{k} \times \langle \psi_\alpha^F(\mathbf{k}) | \nabla_\mathbf{k} | \psi_\alpha^F(\mathbf{k}) \rangle \tag{15}$$

with $\alpha = u, d$. Here, $\psi_\alpha^F(\mathbf{k})$ are the Floquet eigenvectors of $H_1(\mathbf{k})$. The associated Chern number is defined as the integration of the Berry curvature over the entire frist Brillouin zone (BZ),

$$C_\alpha^F = \frac{1}{2\pi} \int_{BZ} d^2\mathbf{k}\, \Omega_\alpha^F(\mathbf{k}) \tag{16}$$

To investigate topological properties of interacting Floquet magnon at finite temperatures, we can put the Floquet magnon band physics on one Bloch sphere, and topological natures are included in one total mass term $m_\mathbf{k} = 2D_F S \rho_\mathbf{k} - \delta_\mathbf{k}$. For the convenience of numerical calculation, one can adopt the Berry curvature of pseudospin freedom [66], which is

$$\Omega_\alpha(\mathbf{k}) = \mp \frac{1}{2d^3} \mathbf{d} \cdot \left( \frac{\partial \mathbf{d}}{\partial k_x} \times \frac{\partial \mathbf{d}}{\partial k_y} \right). \tag{17}$$

where, $d = \sqrt{d_x^2 + d_y^2 + d_z^2}$, and $\mathbf{d} = (d_x, d_y, d_z)$ is the effective field vector of Eq.(14). By manipulating the temperature-dependent Floquet magnon population, we can achieve a topological phase transition in the upper band of the renormalized

Floquet magnon from $C_u^F = -1$ to $C_u^F = 1$.

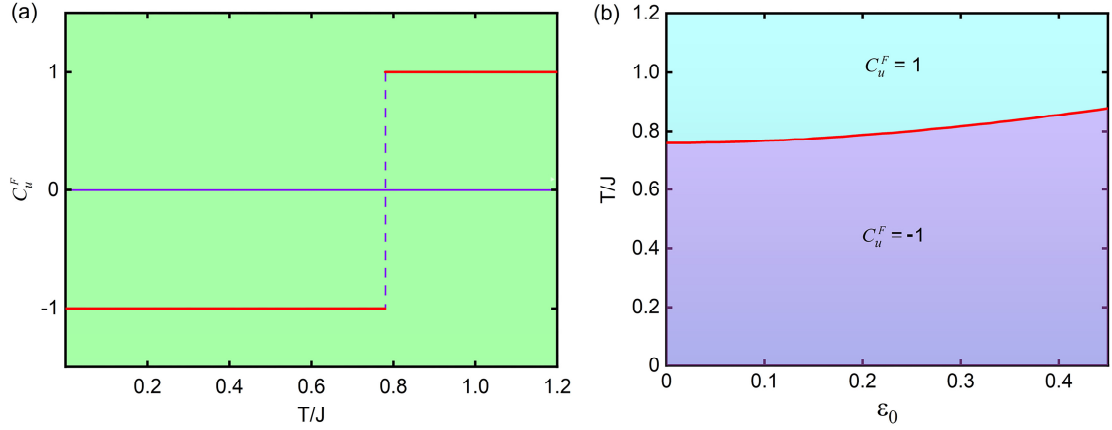

**Fig.6.** (Color online) (a)The Chern number of the upper band of the renormalized Floquet magnon, with $\varepsilon_0 = 0.15$. (b) The phase diagram in the $\varepsilon_0 - T$ plane for the upper band of the renormalized Floquet magnon, and the Chern number of the down band of the renormalized Floquet magnon is the opposite. The other parameters are chosen as $D = 0.1J$, $h = 0.1J$, and $S = 1/2$.

In Fig. 6, we display the dependence of the Chern number for the upper band of the renormalized Floquet magnon on the temperature and the phase diagram in the plane for the upper band of the renormalized Floquet magnon. There is a topological phase transition driven by temperature, the Chern number for the upper band changes below $T_c$ and above $T_c$ as shown in Fig. 6(a), and the renormalized magnon gaps of the Dirac points $K_+$ close and reopen near the critical temperature $T_c$. The closing of band gap at the transition point is essential to ensure the topological phase transition [7]. To further investigate temperature-induced topological transitions at the tunable light intensity $\varepsilon_0$, we plot the phase diagram in the $\varepsilon_0 - T$ plane for the upper band of the renormalized Floquet magnon as shown in Fig. 6(b). The results show that the critical temperature $T_c$ of temperature-induced topological phase transition increases with the increase of light intensity. It is easy to understand that with the application of laser field, the appearance of photo-induced magnetic order contributes to the stability of ferromagnetic phase, which means that phase transition requires a higher temperature. More generally, in Heisenberg ferromagnets, the exchange interaction between magnetic atoms is one of the main causes of ferromagnetism. The light intensity can affect the exchange interaction and thus the magnetism of the ferromagnet. When the light intensity increases, the exchange interaction may be enhanced, resulting in enhanced magnetism of the ferromagnet. In addition, light intensity can also affect the magnetism of ferromagnets by affecting the thermal motion and lattice vibration of the material. With the increase of light intensity, the thermal motion and lattice vibration of the material may be enhanced, resulting in changes in magnetic properties. In summary, the light intensity has a complex effect on the magnetism of ferromagnets. At a certain temperature, with the increase of light intensity, the enhancement of exchange interaction, thermal motion and lattice vibration may cause the magnetic properties of ferromagnets to change, thus affecting the critical temperature of the topological phase transition induced by temperature.

## VI. THERMAL HALL EFFECT OF INTERACTING FLOQUET MAGNON

The most interesting aspects of ferromagnetic topological magnons is that they exhibit the thermal Hall effect. Topological magnon systems, such as the magnonic analog of spin Hall insulators and Weyl semimetals, have been extensively explored in recent years. These systems host linear magnon spin Nernst or thermal Hall currents, which are induced by non-collinear spin texture or DMI. However, in the absence of DMI the linear thermal Hall signal vanishes and transport signatures in the Hall response appear only in the nonlinear response regime. This has motivated the recent exploration of exciting nonlinear transport phenomena in bosonic systems, especially the nonlinear thermal Hall effect. In ferromagnetic insulators, the thermal Hall effect has only been studied in the topological magnon insulator phase, when the lowest (acoustic) magnon band is well separated and carry a well-defined Chern number. For periodically driven magnon systems, we focus on the regime where the

Bose distribution function is close to thermal equilibrium. In this regime, the same theoretical concept of the thermal Hall effect in undriven topological magnon systems can be applied to the driven magnon systems. The transverse component of the thermal Hall conductivity [26, 27] is given explicitly by

$$\kappa_{xy} = -\frac{k_B^2 T}{V} \sum_{\mathbf{k}} \sum_{\alpha=u,d} c_2(n_\alpha) \Omega_\alpha(\mathbf{k}) \qquad (18)$$

where $V$ is the volume of the system, $k_B$ and $T$ are the Boltzmann constant and the temperature respectively. And $n_{\mathbf{k},\alpha} = f(\varepsilon_\alpha(\mathbf{k})) = \left[e^{\beta \varepsilon_\alpha(\mathbf{k})} - 1\right]^{-1}$ corresponds to well-known Bose-Einstein distribution function close to thermal equilibrium, $\beta = 1/k_B T$, $c_2(x) = (1+x)\left(\ln\frac{1+x}{x}\right)^2 - (\ln x)^2 - 2\text{Li}_2(-x)$, and $\text{Li}_2(x)$ stands for the dilogarithm function.

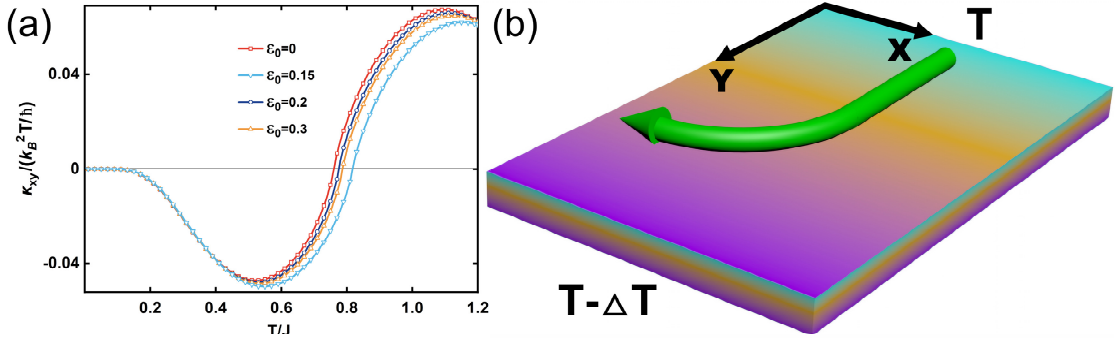

**Fig.7.** (Color online) (a) The thermal Hall conductivity as a function of temperature at different light intensities. (b) Schematic figure of the thermal Hall effect. The parameters are chosen as $D = 0.1J$, $h = 0.1J$, and $S = 1/2$.

Evidently, the thermal Hall conductivity is simply the Berry curvature weighed by the $c_2$ function. Therefore, its dominant contribution comes from the peaks of the Berry curvature. In addition, the thermal Hall transport is determined mainly by the acoustic (lower) magnon branch due to the bosonic nature of magnons. In Fig. 7(a), we have shown the trend of the thermal Hall conductivity $\kappa_{xy}$ for different light intensities as the temperature increase. The main result of this report is that intrinsic topological magnon insulators in the honeycomb ferromagnets can be driven to different topological phases with different Berry curvatures using photo-irradiation. Therefore, each topological phase is associated with a different sign of the thermal Hall conductivity, which results in a sign reversal of the magnon heat photocurrent. The finite thermal Hall conductivity in these topologically trivial phases can be attributed to the photoinduced Berry curvatures. We all know that the thermal Hall conductivity vanishes for the magnonic Floquet trivial insulator induced by periodically circularly polarized light $\varphi = \pi/2$ without regard to the magnon-magnon interaction, because the integration of the Berry curvature vanishes. But for Interacting topological Dirac magnons system, the thermal Hall conductivity always exists at a small light intensity as the Berry curvature exists. In other words, the thermal Hall effect in 2D ferromagnetic insulators is not necessarily a consequence of topological magnon insulator, but it depends solely on the Berry curvature of the magnon bands. This is expected as periodically circularly polarized light will break timereversal symmetry.

As shown in Fig.7, the magnon-magnon interaction has an important influence on the thermal Hall effect in Heisenberg honeycomb ferromagnets. This interaction can occur in two ways: exchange interaction and direct interaction. Exchange interaction is the main mode of interaction between magnons in ferromagnets. In this interaction, the spin directions of magnons will influence each other, resulting in a change in the orientation of the spin magnetic moments. This change in orientation affects the thermal Hall effect because the heat flow must overcome the exchange interaction between the spin magnetic moments as it passes through the ferromagnet, resulting in a reduction in the heat flow. Therefore, the presence of exchange interactions results in a positive sign for the thermal Hall conductivity. Direct interaction is another way of interaction between magnons, which can occur through the mechanism of many-body interaction. In this interaction, the spin directions of magnons will influence each other, resulting in a change in the distribution of spin magnetic moments. This change in distribution affects the thermal Hall effect because the heat flow must pass through the distribution region of the spin magnetic moment when passing through the ferromagnet. When the distribution of spin magnetic moments changes, the transmission mode of heat flow will also change, resulting in a change in the sign of the thermal Hall conductivity. In addition, as the temperature increases, thermal excitation will gradually destroy the ordered arrangement between magnons, resulting in a change in the interaction between magnons. This change may cause the sign of the thermal Hall conductivity to reverse. Thus, in a Heisenberg honeycomb ferromagnet, the influence of magnon-magnon interaction on the thermal Hall effect may change with temperature as our study shown.

## VII. SUMMARY AND OUTLOOK

In summary, we have investigated the many-body interaction effects of Floquet magnons in laser-irradiated Heisenberg honeycomb ferromagnet with off-resonant circularly polarized light field tunable DMI. Our results show that the quantum ferromagnet systems periodically laser-driven exhibits temperature-driven topological phase transitions due to the interaction between Floquet magnons. These transitions are marked by the Floquet magnon band gap closing-reopening with increasing temperature and the sign change of the Chern number and thermal Hall conductivity with increasing temperature at a finite off-resonant circularly polarized light field and magnetic field. Our work complements the lack of previous theoretical works that the topological phase transition of the Floquet magnon under the linear spin wave approximation is only tunable by the light field. The thermal Hall effect of Floquet magnons provides a prominent signature of the topological phase transitions near the point $K_+$ - allowing for the experimental investigation of this many-body effect. Our proposal and conclusion are quite universal for quantum ferromagnet systems driven periodically and can be extended to quantum antiferromagnet systems driven periodically. In general, we believe that the results in this paper are pertinent to experiments and will remarkably impact future research in topological magnon insulators, topological insulating antiferromagnets and their potential practical applications to photo-magnonics and magnon spintronics. In experiments, the Cr-based trihalide $CrX_3$ (X=F, Cl, Br, and I) may be viewed as a potential candidate to confirm our results.

## ACKNOWLEDGMENTS

This work was supported by the National Natural Science Foundation of China under Grant No. 12064011 and the Scientific Research Fund of Hunan Provincial Education Department under Grant No. 23A0404.